\input{aipcheck}

\documentclass[
    ,final            
  ]
  {aipproc}

\layoutstyle{8x11double}

\begin{document}

\title{Neutrinoless double beta decay search\\ with the NEMO~3 experiment}


\author{Irina Nasteva on behalf of the NEMO collaboration}{
  address={Particle Physics Group, School of Physics and Astronomy, University of Manchester, Manchester, M13 9PL, UK}, email={Irina.Nasteva@manchester.ac.uk}}

\begin{abstract}
The NEMO~3 experiment searches for neutrinoless double beta decay and makes precision measurements of two-neutrino double beta decay in seven isotopes. 
The latest two-neutrino half-life results are presented, together with the limits on neutrinoless half-lives and the corresponding effective Majorana neutrino masses. 
Also given are the limits obtained on neutrinoless double beta decay mediated by $R_p$-violating SUSY, right-hand currents and different Majoron emission modes. 
\end{abstract}

\maketitle


\section{Introduction}

Neutrinoless double beta decay ($0\nu \beta \beta$) is a process in which two neutrons in a nucleus undergo simultaneous beta decays with the emission of two electrons.
It violates lepton number and is therefore forbidden in the Standard Model.
The observation of neutrinoless double beta decay would prove that neutrinos are Majorana particles and would provide access to their absolute mass scale.

The half-life of $0\nu \beta \beta$ is given by:
\begin{equation}
\frac{1}{T^{0\nu}_{1/2}(A,Z)}=|M^{0\nu}(A,Z)|^2 G^{0\nu}(Q,Z) \langle m_{\beta\beta} \rangle ^2\,,
\label{eq1}
\end{equation}
where $M^{0\nu}(A,Z)$ is the nuclear matrix element (NME) and $G^{0\nu}(Q,Z)$ is a known phase
space factor that depends on the transition energy ($Q$-value) of the process. 
The effective Majorana neutrino mass, $\langle m_{\beta\beta} \rangle$, is a sum over the mass eigenstates, weighted by the squared elements, $U^2_{ei}$, of the PMNS neutrino mixing matrix: 
\begin{equation}
\langle m_{\beta\beta} \rangle=\sum_{i=1,2,3} U_{ei}^2 m_i \,.
\label{eq2}
\end{equation}

In addition to light Majorana neutrino exchange (described by \eqref{eq1}), $0\nu \beta \beta$ decay could be mediated by other physics mechanisms such as a right-hand current admixture in the Lagrangian, Majoron emission or supersymmetric particle exchange. 
Measuring the energy spectrum and topology of the final state electrons could allow to distinguish between these underlying mechanisms. 

\section{The NEMO~3 experiment}

NEMO~3 \cite{nemo3tdr} (Neutrino Ettore Majorana Observatory) has been taking data since 2003 at the Modane Underground Laboratory in France.
The experiment is dedicated to searching for $0\nu \beta \beta$ decay and making precise lifetime measurements of $2\nu \beta \beta$ decay in seven isotopes.
The main isotopes used for the $0\nu \beta \beta$ search are 7~kg of $^{100}$Mo and 1~kg of $^{82}$Se. 
There are smaller amounts of $^{116}$Cd, $^{130}$Te, $^{150}$Nd, $^{96}$Zr and $^{48}$Ca for $2\nu \beta \beta$ studies, and natural Te and Cu for background measurements.

NEMO~3 employs an experimental technique of calorimetry and tracking, in order to detect the two final state electrons.
The detector is cylindrical and is radially segmented into 20 equal sectors, 
each housing a thin source foil placed in the middle of a tracking volume, which is surrounded by the calorimeter.
The tracker consists of 6180 drift cells operated in Geiger mode.
The calorimeter walls are made up of 1940 plastic scintillator blocks coupled to low-radioactivity PMTs. They achieve energy resolution in the range 14\%--17\% FWHM for 1~MeV electrons and timing resolution of 250~ps.
A solenoid magnetic field of 25~G is applied to provide charge identification.
The whole detector is enclosed in a radon-free air tent, and covered by two levels of external radiation shielding.

The NEMO~3 detector measures the individual particle trajectories and energies, thus reconstructing the final state topology and kinematics of the events.
Through particle identification of $e^-$, $e^+$, $\alpha$ and $\gamma$ it achieves excellent background suppression, which is further enhanced by the time-of-flight measurement used to reject external particles crossing the detector. 

Double beta decay events are selected by requiring two tracks with a negative curvature, originating from a common vertex in the source foil and associated to isolated scintillator energy deposits. 
The timing of the calorimeter hits is required to agree with the time-of-flight hypothesis of two electrons emitted from the foil at the same time. 
The remaining backgrounds in the two-electron signal sample are estimated by looking at control channels.

\section{$2\nu\beta\beta$ results}

Two-neutrino double beta decay ($2\nu \beta \beta$) is a Standard Model weak interaction process 
occurring in nuclei for which beta decay is energetically forbidden or strongly suppressed. 
The importance of understanding the $2\nu\beta\beta$ process is because it forms the irreducible background to $0\nu\beta\beta$ decay. 
In addition, precise measurements of its half-life and event kinematics are used to constrain the nuclear models used to calculate the neutrinoless NME.

\begin{figure}
  \includegraphics[height=.3\textheight]{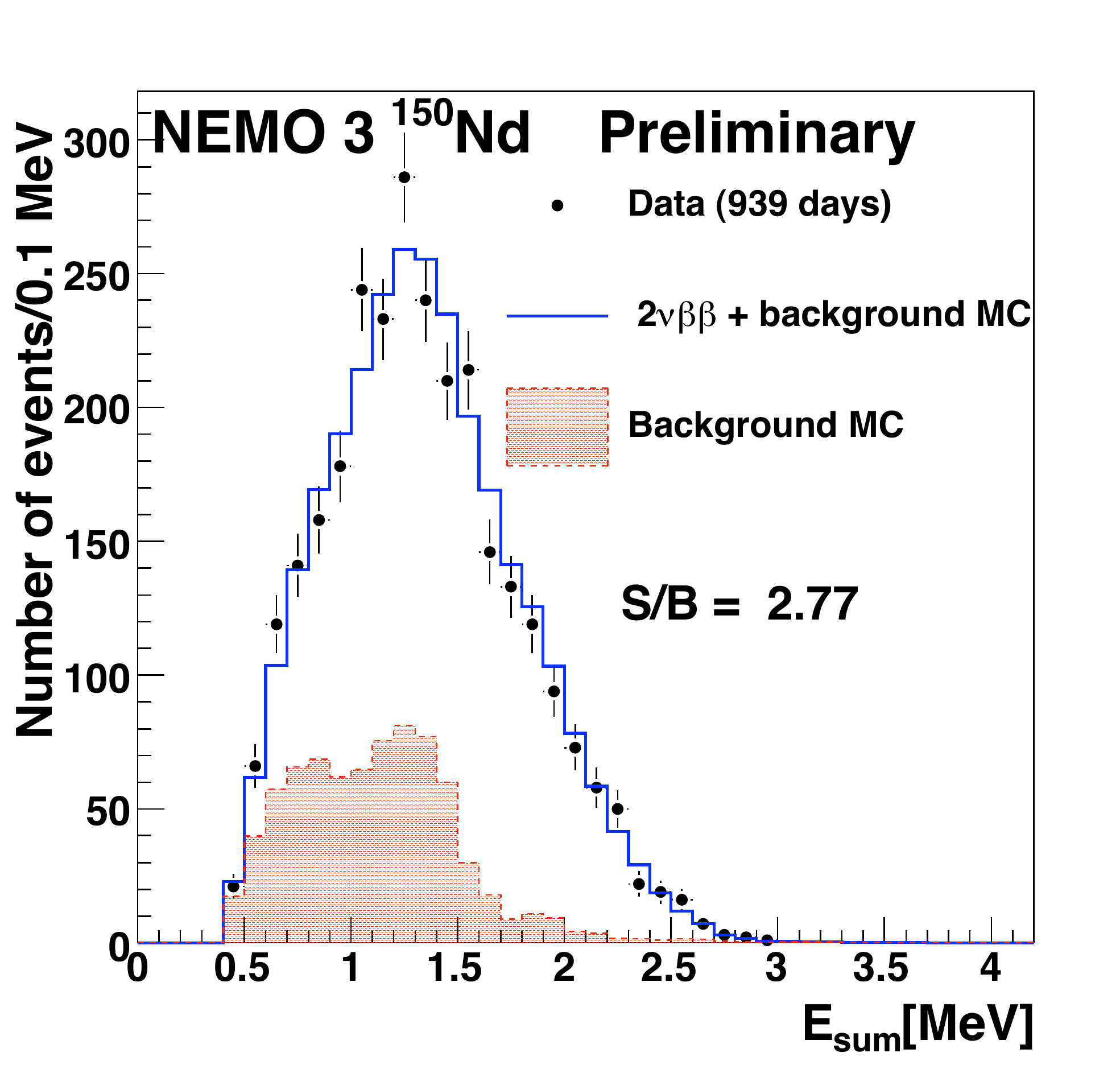}
  \caption{Energy sum distribution of the two electrons observed in $^{150}$Nd decays.}
  \label{ndsum}
\end{figure}

\begin{figure}
  \includegraphics[height=.23\textheight]{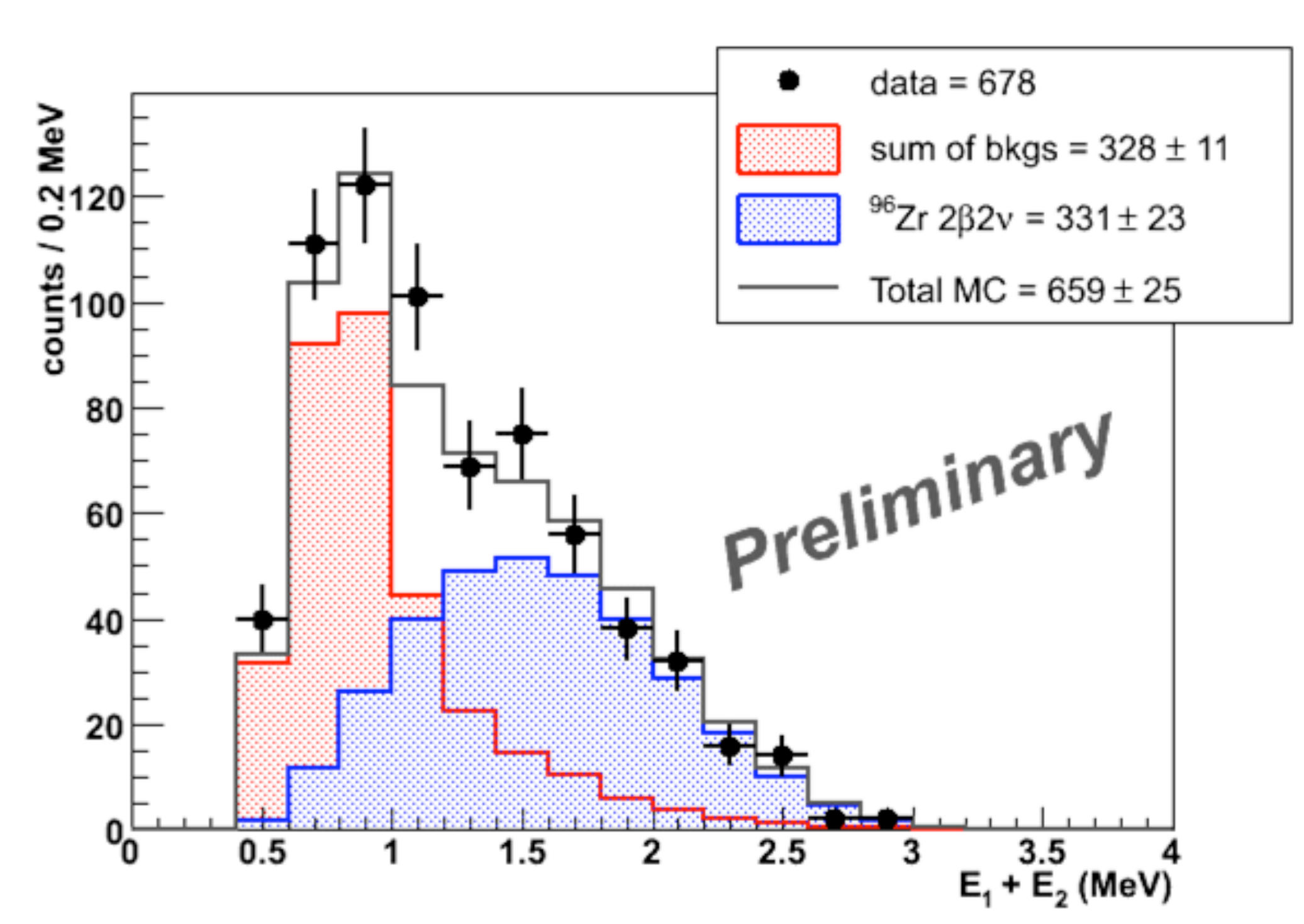}
  \caption{Energy sum distribution of the two electrons observed in $^{96}$Zr decays.}
  \label{zrsum}
\end{figure}

The NEMO~3 experiment has been performing high-statistics measurements of $2\nu\beta\beta$ decay in its seven isotopes. 
Figures \ref{ndsum} and \ref{zrsum} show the recent preliminary results for the two-electron energy sum distributions obtained from $^{150}$Nd and $^{96}$Zr, respectively.
Table \ref{all2nu} summarises the current half-life measurements, along with the isotope characteristics and the signal-to-background ratios (S/B), obtained from all isotopes.

\begin{table}
\begin{tabular}{lcccc}
\hline
\tablehead{1}{r}{b}{Isotope} 
  & \tablehead{1}{c}{b}{Mass (g)}
  & \tablehead{1}{c}{b}{$Q$ (MeV)}
  & \tablehead{1}{c}{b}{S/B}
  & \tablehead{1}{c}{b}{$T_{1/2}^{2\nu\beta\beta}$ ($\times 10^{19}$ y)} \\
\hline
$\rm ^{100}Mo$ &6914& 3.034 &40 & 0.711 $\rm\pm$ 0.002 (stat) $\pm$ 0.054 (syst)~\cite{nemoresult1}\\
$\rm ^{82}Se$  &932 & 2.995 & 4   & 9.6 $\pm$ 0.3 (stat) $\pm$ 1.0 (syst)~\cite{nemoresult1}\\
$\rm ^{130}Te$ &454 & 2.529& 0.25  & 76 $\pm$ 15 (stat) $\pm$ 8 (syst)~\cite{ssr}\\
$\rm ^{116}Cd$ & 405& 2.805 & 7.5 & 2.8 $\pm$ 0.1 (stat) $\pm$ 0.3 (syst)~\cite{ssr}\\
$\rm ^{150}Nd$ &37.0& 3.367& 2.8  & $0.920 ^{+0.025}_{-0.022} $(stat) $\pm$ 0.073 (syst)\\
$\rm ^{96}Zr$  &9.4 & 3.350 & 1.0 & 2.3 $\pm$ 0.2 (stat) $\pm$ 0.3 (syst)\\
$\rm ^{48}Ca$  &7.0 & 4.272 & 6.8  & $4.4 ^{+0.5}_{-0.4} $(stat) $\pm$ 0.4 (syst)\\
\hline
\end{tabular}
\caption{NEMO~3 results for $2\nu\beta\beta$ half-life measurements for seven isotopes.}
\label{all2nu}
\end{table}

\section{$0\nu \beta \beta$ search}

\subsection{Light neutrino exchange}

\begin{figure}
  \includegraphics[height=.28\textheight]{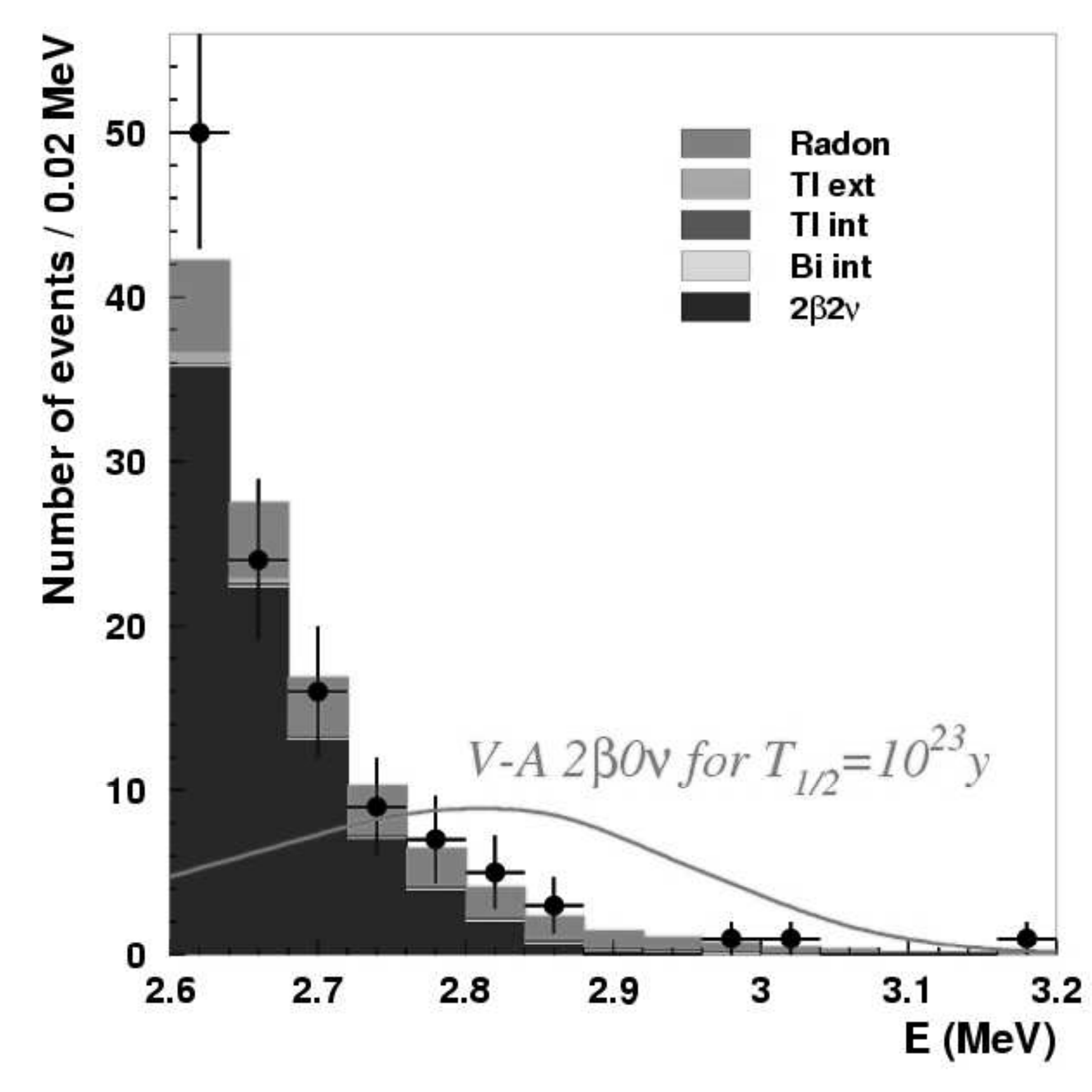}
  \caption{The endpoint of the $^{100}$Mo two-electron energy sum distribution. The line shows a simulation of a neutrinoless signal of half-life $10^{23}$ years.}
  \label{mosum}
\end{figure}

\begin{table}
\begin{tabular}{lcc}
\hline
\tablehead{1}{r}{b}{Isotope} 
  & \tablehead{1}{c}{b}{$T_{1/2}^{0\nu\beta\beta}$ (y)} 
 & \tablehead{1}{c}{b}{$\langle m_{\beta\beta} \rangle$ (eV)} \\
\hline
$\rm ^{100}Mo$ & $>5.8\times 10^{23}$& $< 0.6-1.3$ \cite{kort75,kort76,Rodin,Simkovic}\\
$\rm ^{82}Se$  & $>2.1\times 10^{23}$ & $<1.2-2.2$ \cite{kort75,kort76,Rodin,Simkovic}\\
$\rm ^{150}Nd$ & $>1.8\times 10^{22}$& $<1.7-2.4$ \cite{Rodin}\\
& & $<4.8-7.6$ \cite{Hirsch} \\
$\rm ^{96}Zr$  & $>8.6\times 10^{21}$ & $<7.4-20.1$ \cite{kort75,kort76,Rodin,Simkovic} \\
$\rm ^{48}Ca$  & $>1.3\times 10^{22}$ & $<29.6$ \cite{Caurier} \\
\hline
\end{tabular}
\caption{NEMO~3 limits at 90\% CL on the half-lives of $0\nu\beta\beta$ and the corresponding effective neutrino mass ranges.}
\label{all0nu}
\end{table}

Neutrinoless double beta decay mediated by light Majorana neutrino exchange would lead to a peak at the endpoint energy $Q$ of the two electrons, smeared by the energy resolution of the detector. 
The endpoint of the energy distribution of $2\nu\beta\beta$ decay for $^{100}$Mo is shown on Fig.~\ref{mosum}. 
No excess of events was observed, therefore a lower limit on the $0\nu\beta\beta$ half-life was obtained, $T_{1/2}^{0\nu\beta\beta}>5.8\times 10^{23}$~years (90\%~CL). 
This translates into a range of upper limits on the effective Majorana neutrino mass, $\langle m_{\beta\beta} \rangle< 0.6-1.3$~eV, according to the most recent NME calculations~\cite{kort75,kort76,Rodin,Simkovic}. 
Table~\ref{all0nu} summarises the half-life and neutrino mass limits obtained by NEMO~3.

The $0\nu\beta\beta$ analysis of $^{100}$Mo and $^{82}$Se has now been blinded. 
The projected half-life sensitivities when unblinded in 2010 are $2\times 10^{24}$ years for $^{100}$Mo and $8\times 10^{23}$ years for $^{82}$Se, leading to neutrino mass reaches of $0.3-0.7$~eV and $0.6-1.1$~eV, respectively. 

\subsection{SUSY particle exchange}

Neutrinoless double beta decay could be mediated by the exchange of superparticles in $R_p$-violating SUSY~\cite{rpvsusy}. 
The half-life of this process is inversely proportional to the SUSY lepton-number violating parameter $\eta^{11}_{(q)LR}$, which is related to the sum of $R_p$-violating trilinear couplings, $\lambda'_{11k}\lambda'_{1k1}$ ($k=1,2,3$). 
From the $^{100}$Mo half-life limit shown in Table~\ref{all0nu}, a limit was obtained of  $\eta^{11}_{(q)LR}<9.2\times 10^{-9}$ that corresponds to upper limits on the trilinear couplings of $\lambda'_{111}\lambda'_{111}<1.7\times10^{-5}$, $\lambda'_{112}\lambda'_{121}<8.7\times10^{-7}$ and $\lambda'_{113}\lambda'_{131}<3.6\times10^{-8}$~\cite{rpvsusy}.

\subsection{Other exotic mechanisms}

\begin{table}
\begin{tabular}{c c c c c c}
\hline
\tablehead{1}{c}{b}{Mechanism} & \tablehead{1}{c}{b}{$^{100}$Mo $T_{1/2}$ (y)} &  \tablehead{1}{c}{b}{$^{82}$Se $T_{1/2}$ (y)} \\
\hline
(V+A) current & $>3.2\times {10^{23}} \tablenote{$\lambda < 1.8\times 10^{-6}$}$& $>1.2\times 10^{23}$&\\
 $n=1$&  $>2.7\times {10^{22}} \tablenote{$g < (0.4-1.8)\times 10^{-4}$}$& $>1.5\times 10^{22} $\\
 $n=2$&$>1.7\times 10^{22}$&$>6.0\times 10^{21}$\\
 $n=3$&$>1.0\times 10^{22}$&$>3.1\times 10^{21}$\\
 $n=7$&$>7.0\times 10^{19}$&$>5.0\times 10^{20}$\\
\hline
\end{tabular}
\caption{\label{exotic}
Constraints at 90\% CL from NEMO~3 data on the half-lives of exotic
processes, and on the (V+A) Lagrangian parameter $\lambda$ and the Majoron to neutrino coupling strength $g$ \cite{nemoresult2}.}
\end{table}

Other exotic mechanisms such as right-hand ($V+A$) currents and Majoron emission could also contribute to $0\nu\beta\beta$ decay. 
They would lead to a distortion of the shape of the two-electron energy sum distribution. 
A maximum likelihood analysis of the deviation of the energy shape from the calculated $2\nu\beta\beta$ shape was performed~\cite{nemoresult2}. 
The resulting limits on the half-lives and couplings for ($V+A$) currents and different Majoron spectral indices $n$ are shown in Table~\ref{exotic}.

\section{The SuperNEMO project}

The next-generation $0\nu\beta\beta$ project SuperNEMO aims to extrapolate the successful NEMO~3 experimental technique to a detector with $\sim100$~kg of source isotopes. 
It will use the technology of calorimetry and tracking in a modular structure, whilst improving critical performance parameters such as energy resolution, acceptance and source purity. 
SuperNEMO aims to achieve a $0\nu\beta\beta$ half-life sensitivity of $T_{1/2}^{0\nu\beta\beta}>2\times 10^{26}$~years, corresponding to an effective neutrino mass reach of $\langle m_{\beta\beta} \rangle< 0.05-0.1$~eV.





\bibliographystyle{aipproc}   

\bibliography{supernemo}

\IfFileExists{\jobname.bbl}{}
 {\typeout{}
  \typeout{******************************************}
  \typeout{** Please run "bibtex \jobname" to optain}
  \typeout{** the bibliography and then re-run LaTeX}
  \typeout{** twice to fix the references!}
  \typeout{******************************************}
  \typeout{}
 }

\end{document}